\documentclass[a4paper,english]{article}
\usepackage[T1]{fontenc}
\usepackage[latin9]{inputenc}
\usepackage{color}
\usepackage{float}
\usepackage{textcomp}
\usepackage{amstext}
\usepackage{graphicx}
\usepackage{apalike}

\usepackage[final]{changes}
\colorlet{Changes@Color}{red}

\makeatletter

\newcommand{\lyxaddress}[1]{
\par {\raggedright #1
\vspace{1.4em}
\noindent\par}
}

\date{}
\makeatother

\usepackage{babel}
\begin{document}

\title{Automated image acquisition for low-dose STEM at atomic resolution}
\maketitle
\begin{center}
Andreas Mittelberger, Christian Kramberger, Christoph Hofer, Clemens
Mangler, Jannik C. Meyer{*}
\par\end{center}

\lyxaddress{\begin{center}
University of Vienna, Faculty of Physics, Boltzmanngasse 5, 1090
Vienna, Austria
\par\end{center}}

{*}corresponding author: jannik.meyer@univie.ac.at
\begin{abstract}
Beam damage is a major limitation in electron microscopy that becomes
increasingly severe at higher resolution. One possible route to circumvent
radiation damage, which forms the basis for single-particle electron
microscopy and related techniques, is to distribute the dose over
many identical copies of an object. For the acquisition of low-dose
data, ideally no dose should be applied to the region of interest
prior to the acquisition of data. We present an automated approach
that can collect large amounts of data efficiently by acquiring images
in an user-defined area-of-interest with atomic resolution. We demonstrate
that the stage mechanics of the Nion UltraSTEM, combined with an intelligent
algorithm to move the sample, allows the automated acquisition of
atomically resolved images from micron-sized areas of a graphene substrate.
Moving the sample stage automatically in a regular pattern over the
area-of-interest enables the collection of data from pristine sample
regions without exposing them to the electron beam before recording
an image. Therefore, it is possible to obtain data with \replaced{minimal} {arbitrarily
low} dose (no prior exposure from focusing),\added{ which is only limited by the minimum signal needed for data processing}\replaced{. This enables us}{, in order} to prevent beam induced damage in the sample and to acquire large datasets within a reasonable amount of time.
\end{abstract}

\section*{Introduction}

Automating data acquisition tasks in transmission electron microscopy
(TEM) started with the availability of CCD cameras and computer-controlled
TEMs back in the early 90's \cite{Dierksen1992,Koster1992}. One of
the main goals behind automation was, and still is \cite{Mastronarde2005},
reducing the dose needed in electron tomography as well as making
tilt series acquisition less prone to errors introduced by manual
re-centering and re-focusing. Another area where automated data acquisition
is important is nowadays in single-particle electron microscopy \cite{Suloway2005,Shi2008,Zhang2001,Zhang2003,Zhang2010a,Cheng2015a,Bartesaghi2015,Frank1978}.
This technique uses low-dose images of many copies of the same object
to reduce beam damage in highly radiation-sensitive specimens such
as biological samples. After collecting hundreds to thousands of images
containing a total of up to a few ten thousand particles, the dataset
has to be aligned and classified. Classification means that particle
images are divided into groups with similar orientation and conformation
which makes it possible to average them in the next step and therefore
to greatly increase the signal-to-noise ratio. The resulting dataset
consists now of different two-dimensional projections of the imaged
particles which can be used to calculate a three-dimensional model
in the next step \cite{Cheng2015}.

Besides using small doses for the actual image acquisition, ``low-dose''
also means separating the focusing and sample-tracking part from the
actual recording of image data. In the ideal case, every snapshot
should be recorded from a ``fresh'' region of the sample, i.e.,
no dose should be applied to the image area for focusing and the image
acquisition should start with the first electron in the respective
field of view. Our approach is closely related to automated methods
used in single-particle analysis with the difference that we aim for
much smaller structures, and atomic resolution. Our test sample is
a graphene sheet suspended over the holes of a Quantifoil(TM) grid,
which serves very well for judging the alignment and focus precision.
Directly conceivable applications would be the imaging of radiation-sensitive
defects in 2D materials, or the analysis of small molecules deposited
on the graphene support or embedded in a graphene sandwich. However,
applications to other ultra-thin samples, such as molecules embedded
in thin films also do not seem far-fetched as long as the membranes
are sufficiently flat and thin. This type of data acquisition is also
relevant for our recently published algorithm for structure recovery
from low-dose (S)TEM images, which requires up to a few thousand images
for a successful reconstruction \cite{Meyer2014,Kramberger2016}.
We further demonstrate that we can reconstruct defects created by
electron beam irradiation in the graphene lattice by feeding low-dose
data acquired with the method described here into this algorithm.

\section*{Materials and Methods}
\subsection*{Overview of the mapping procedure}

\begin{figure}
	\begin{centering}
		\includegraphics[width=0.75\textwidth]{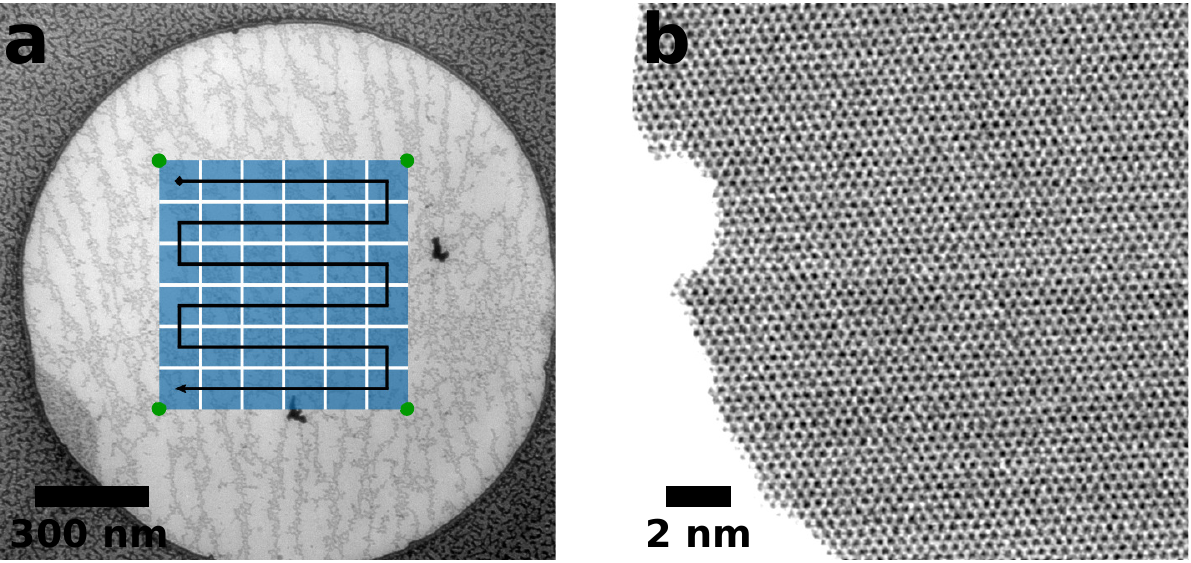}
		\par\end{centering}
	\caption{\label{fig:pattern}(a) Schematic of the approach. Four positions
		are used as focus reference (green dots) and then the central part
		of the membrane is scanned by moving the stage and interpolating the
		focus value. (b) Example of a high-dose image of a graphene membrane,
		recorded by moving to a fresh area automatically and using focus interpolation
		without further adjustment.}
\end{figure}

Electron doses low enough to image for example individual organic
molecules have to be below $10^{0}-10^{3}\,e{}^{-}/\mathring{A}^{2}$
for typical (S)TEM high-voltages \cite{Egerton2012}. In contrast,
atomically resolved images of light elements such as carbon require
doses of at least $10^{5}\,e{}^{-}/\mathring{A^{2}}$. This means
that tuning and focusing have to take place in sample areas other
than the area-of-interest. We achieve this by manually setting a focus
reference on the four corners of a quadrangle that surrounds the area
of interest. This focus references and all further focus corrections
are only set electronically without adjustment of the mechanical height
(z) drive. After the user has set the imaging parameters for the map,
our program fills the quadrangle with a regular array of image positions
and interpolates the focus at each point (see figure \ref{fig:pattern}). Now, the aim is to move
the stage to each position and acquire an image. Remarkably, the deviation
of the stage from the desired in-plane (x,y) coordinates is quite
small (typically below 25 nm for movements within $\sim1\mu\textrm{m}$)
and it appears that stage movements in x- and y-direction do not lead
to any detectable random changes in sample height (z). More importantly,
the deviation in (x,y) from the target position is highly reproducible
if the sequence of (x,y) positions is driven in a specific sequence.
Hence, it becomes possible to measure and compensate the small remaining
inaccuracies of the stage drive. So far, it appears that a ``reference''
measurement for this purpose is needed only once for this sample stage
and (x,y) position pattern. This procedure of course requires the
microscope to provide stable imaging conditions during the whole data
acquisition, which is typically up to a few hours. With the present
parameters we can acquire about 1000 images per hour.

\subsection*{Details of the implementation}

Our microscope is controlled via Nion's own user software, Swift,
which is python-based and also provides an API for custom extensions.
Therefore all programs used for this paper were also written in python
as plug-ins for Nion Swift. In the following paragraph, some details of
the mapping process and its implementation will be explained.

We chose a serpentine-path as mapping pattern e.g. the stage moves
from left to right in odd and from right to left in even lines. The
reason for this is that while the sample stage suffers only very little
from backlash and drift, it is still favorable to avoid large stage movements.
To account for the still present sample drift after moving the stage,
we use a delay of typically 2 s between moving the stage and acquiring
the image. The electron beam is blanked during that time to prevent
additional dose put into the area-of-interest. Moreover, the serpentine
pattern is oriented to be aligned with the natural stage coordinates
(i.e., movement along a line uses only the x or y drive, and not a
combination of x and y). As the four user-selected points usually
form an arbitrary quadrangle, the largest-possible rectangle that
is aligned with the natural stage coordinates is inscribed into them.
This area is then filled with an evenly spaced grid of coordinates,
based on the image size and the chosen offset between the images.
For each stage position the correct focus is calculated via a bilinear
interpolation function. An image is then acquired at each stage position
and saved on hard disk. Also an overview image is taken after the
map has been completed.

\begin{figure}
\centering{}\includegraphics[width=1\textwidth]{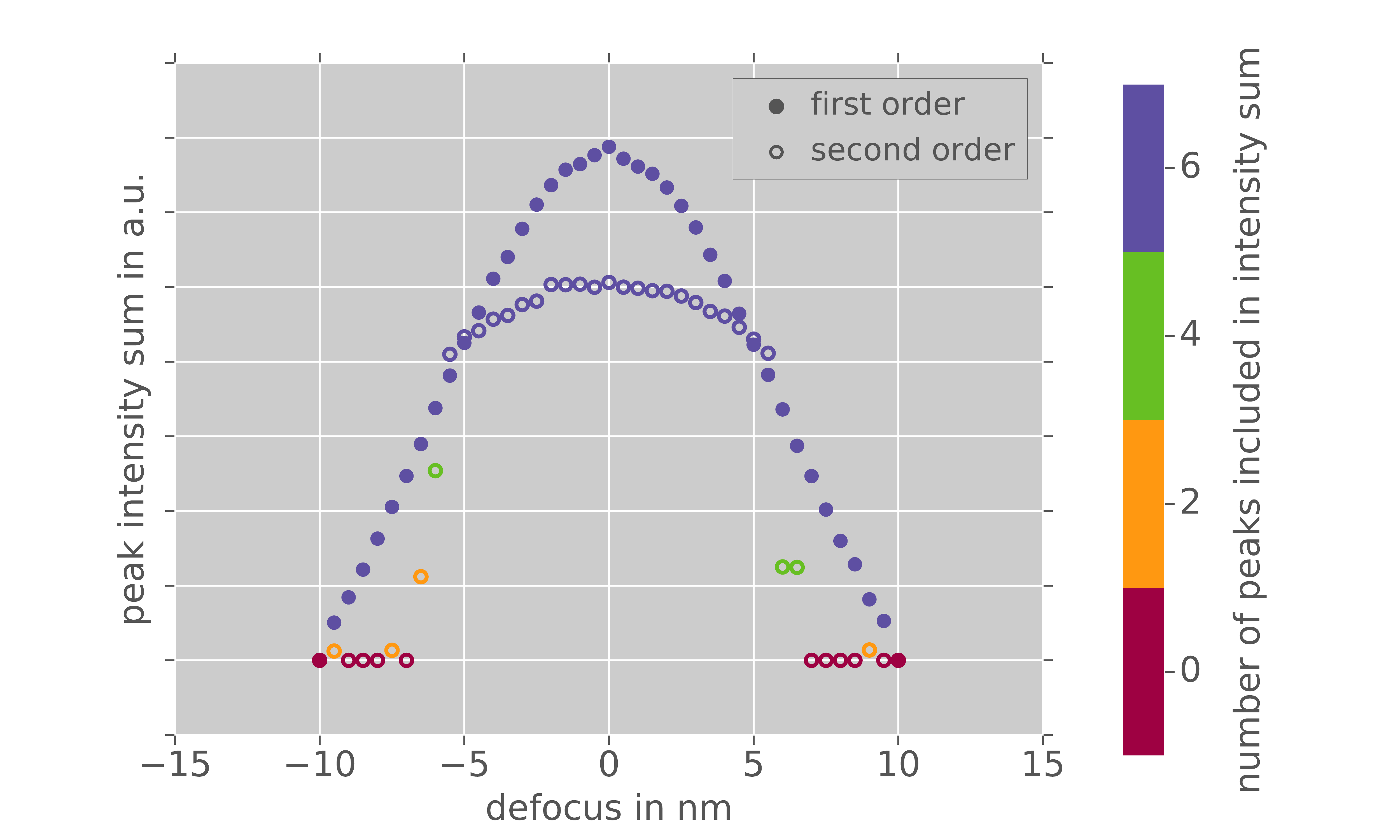}\caption{\label{fig:peak_intensities}Plot of the sum of all visible FT spot
intensities of a simulated ADF-STEM focus series of graphene. The
peak intensity sum is strongly dependent on the amount of defocus
and can therefore be used to judge image quality. Jumps in the curve
occur when the number of peaks included in the intensity sum decreases
(because they become too weak to be found). The number of peaks included
in the sum is visualized by a color code in the image.}
\end{figure}

With this simple approach it is possible to acquire maps consisting
of up to about 1000 images with reasonable quality. As a measure for
judging the focus precision, we use Fourier transforms (FTs) of the
acquired images. For periodic structures like graphene, the peaks
in the FTs are still present even in low-dose exposures where the
lattice is hidden in the noise. For an incoherent (annular dark-field,
ADF) image, the FT peak intensities are a simple but accurate measure
for the quality of atomically resolved images: Figure \ref{fig:peak_intensities}
shows the sum of the first- and second-order peaks of a graphene lattice
in a simulated focus series. Especially the first-order peaks show
a sharp maximum at zero defocus. Because all our samples are either
pure graphene or at least graphene supported, the 12 peaks of the graphene
lattice can be used for judging image quality.

\section*{Results and Discussion}

\begin{figure}
\begin{centering}
\includegraphics[width=1\textwidth]{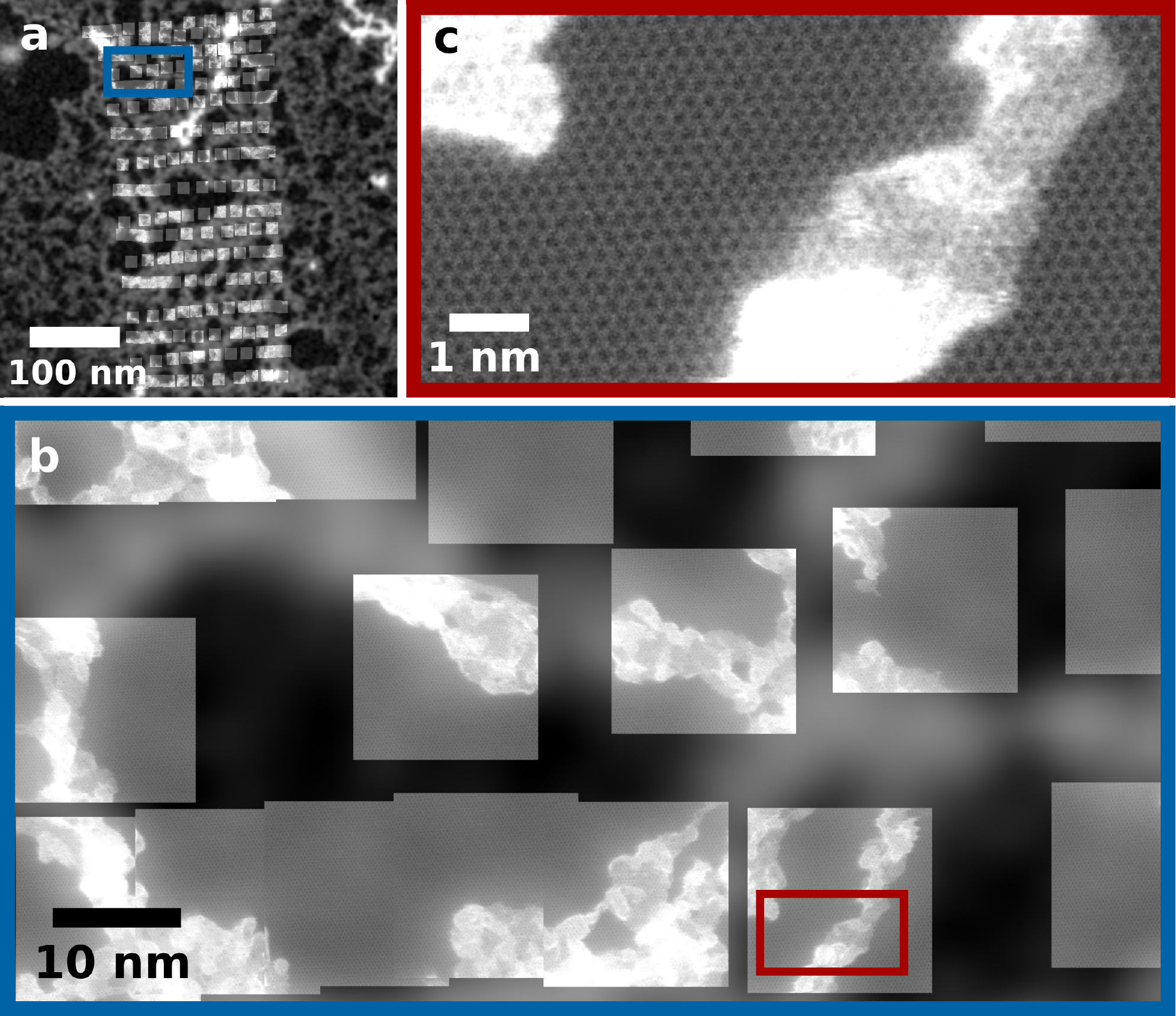}
\par\end{centering}
\caption{\label{fig:high-dose-zoom}Zoom into a part of an automatically acquired
high-dose map consisting of 12x17 images. (a) The atomically resolved
images were aligned on an overview image in order to find their actual
positions on the sample. (b) An about 10 times magnified part of the
map highlighted by the blue rectangle in (a). The contamination pattern
in the background stems from the overview image and was used to align
the small images with respect to the overview. (c) An about 5 times
zoom into the area of (b) marked with a red rectangle reveals the
atomic lattice in the image.}
\end{figure}

To test our approach, we recorded atomically resolved high-dose maps
from graphene. This is no option for beam-sensitive samples, but for
pure graphene, which is stable at 60 kV, it provides a way to verify
that atomic resolution is reached in the single images. In figure
\ref{fig:high-dose-zoom} such a map is shown where all images were
aligned on an overview image. For this purpose, the random contamination
pattern that is typically found on graphene samples provides a perfect
reference structure that is well visible in both the low-magnification
overview and the high-resolution snapshots. Zooming into a small area
of this map reveals the graphene lattice in those images and shows
that we can reach atomic resolution in almost all exposures. In addition
to judging the single image quality, also the accuracy of stage movement
can be seen via this analysis as discussed later in more detail.

\begin{figure}
\begin{centering}
\includegraphics[width=1\textwidth]{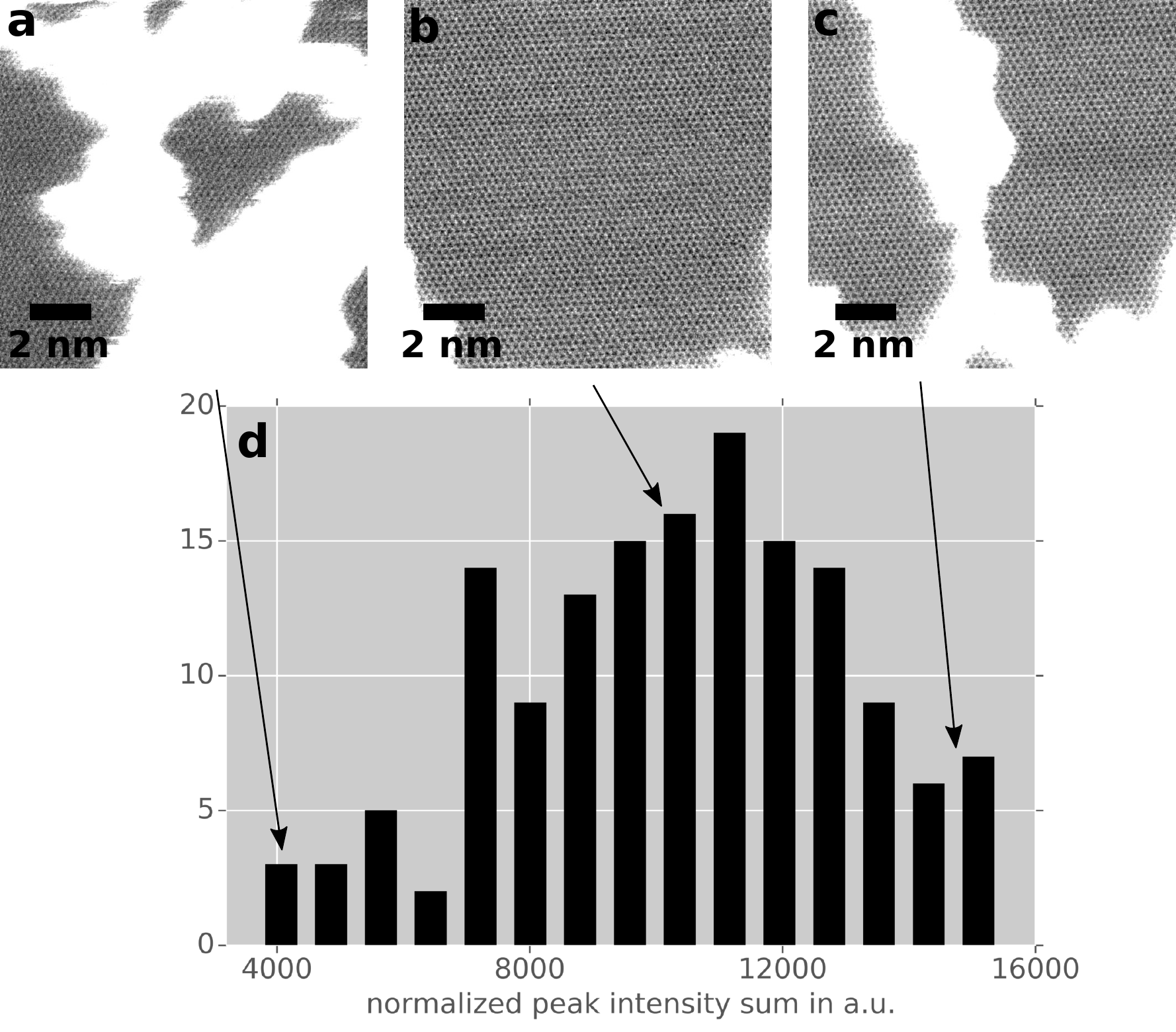}\caption{\label{fig:histogram}Histogram of the peak intensity sums of the
map in figure \ref{fig:high-dose-zoom}. (a) One of the worst focused
images of the map. (b) An image with average focus precision. (c)
One of the best focused images. (d) Histogram of the peak intensity
sums. All peak intensities are normalized with respect to the clean
graphene area in an image to make the intensity sum independent from
contamination coverage.}
\par\end{centering}
\end{figure}

As already mentioned earlier, the FT peak intensity sum in an image
provides a good measure for its tuning. This provides the means to
quantify the overall focus precision in our automatically acquired
maps. Figure \ref{fig:histogram} shows a histogram of all peak intensity
sums of one high-dose map. In panel a, b and c of figure \ref{fig:histogram}
there are three images that correspond to a poorly focused, an averagely
focused and a very well focused image of this map, respectively. According
to the histogram in figure \ref{fig:histogram} d, most images of
a map suffer from slight defocus which is, however, still small enough
to reach atomic resolution (c.f. figure \ref{fig:histogram} b). Assuming
that the best images of a map are acquired at zero defocus, an estimate
for the amount of defocus can be made by assigning a defocus to the
relative peak intensity sum via figure \ref{fig:peak_intensities}.
The maximum peak intensity sum for an image of the map shown in figure
\ref{fig:histogram} is about 16,000 which we assume to be at zero
defocus. Therefore, images with a peak intensity sum of 8,000 (which
is half of the maximum value) are, according to figure \ref{fig:peak_intensities},
recorded with a defocus of $\pm$5 nm. A closer look to such an image
(see figure \ref{fig:histogram} b) still clearly shows the atomic
lattice, indicating that a slight defocus does not significantly decrease
image quality. This finding corresponds well with the expected depth-of-field
for our microscope which is about 10 nm at an acceleration voltage
of 60 kV and an aperture half-angle of 35 mrad \cite{fultz2012transmission}.

\begin{figure}
\begin{centering}
\includegraphics[width=1\textwidth]{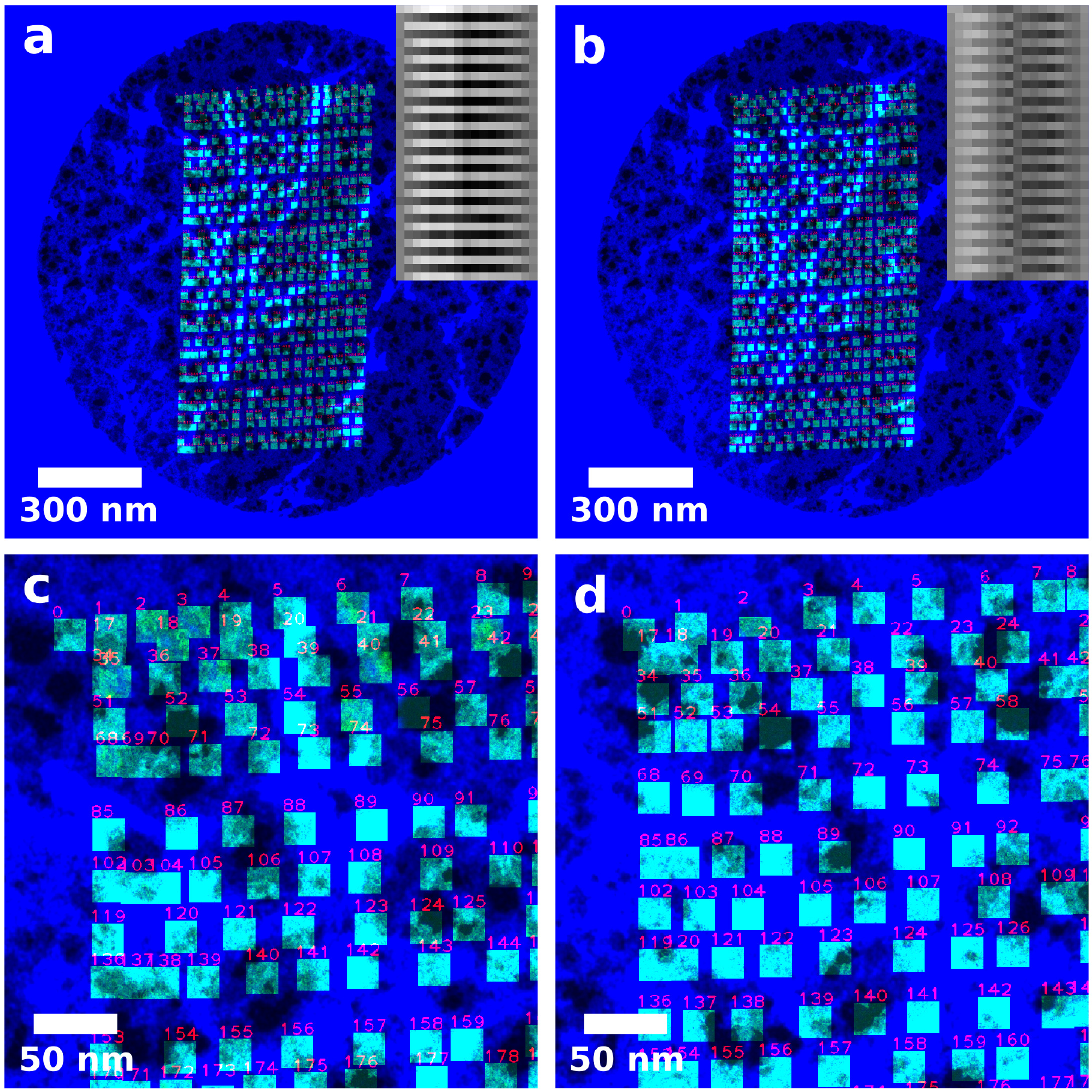}
\par\end{centering}
\raggedright{}\caption{\label{fig:positions}561 automatically acquired images placed on
an overview image to find their actual positions. (a) Frame positions
when feeding a regular grid as stage coordinates into the mapping
program. Inset: Deviation between actual and ideal x-positions as
gray-scale color code. Each square corresponds to one image. The strong
misplacement, especially between even and odd lines is clearly visible.
(b) Another map recorded in the same area as in (a) but with corrected
stage positions. This was done by analyzing frame positions in 10
maps acquired earlier and using the data to create the corrected target
stage positions here. Inset: The x-deviations clearly show the improvement
in comparison with (a). The intensity range was set to $\pm$50 nm
in both insets to make them directly comparable. (c and d) Zoom to
the top-left corner of the map in (a) and (b). The grid of images
in (d) is much more regular than in (c).\added{ Its root mean square displacement from the target positions could be reduced from 36 to 18 nm.}}
\end{figure}

Even though there is a certain focus tolerance it is still desirable
that the actual image positions match the target positions as good
as possible to enable accurate focus interpolation, and to avoid overlapping
exposures. Figure \ref{fig:positions} shows two examples of maps,
where the single frames were aligned on the overview image. To make
it easier to see both, the overview and the single images on top,
the different images were put into the blue and green channel of an
RGB-image, respectively. This allows also to see immediately if a
position was not found correctly by the alignment algorithm because
the colors would not mix to light-blue in this case but the small
image would still appear green. The image numbers are shown in red
in this figure.

The first look at figure \ref{fig:positions}a reveals that the mapped
area is almost a perfect rectangle which matches exactly the target
area. This is due to the fact that the sample stage moves are highly
reproducible, meaning that after going up to 1 \textmu m to the right
and 1 \textmu m back in the next line it arrives again at the left
margin of the rectangle with a precision of a few nm. A closer look,
however, shows that the frames are not distributed in a perfectly
regular pattern although these were the target positions. After the
stage movement direction changed, the first few frames in each line
lie very close to each other and sometimes even overlap because there
is still some backlash present in the stage mechanics. Towards the
end of a line the distance between the images increases. Also in y-direction
there is a periodic deviation from the target positions, although
it is not as pronounced as in x-direction. Another artifact can be
seen in the very first line: The first image is located ca. 30 nm
outside of the rectangle formed by all others and the spacing between
the first and second line is smaller than the distance between following
lines. Some of these artifacts are quantified in the inset in figure
\ref{fig:positions}a, which shows the x-deviation of the actual image
positions from their target positions. Especially the almost-perfectly
aligned borders of the mapped area and the strong deviation (backlash)
between even and odd lines are visible there.

Overlapping areas are not desirable when using this automatic acquisition
procedure for low-dose imaging because beam-sensitive structures would
have already been exposed when the second snapshot is taken. Increasing
the offset between the images is also not a good solution because
it would lead to a big amount of unused sample area. We can significantly
increase the precision of the (x,y) positions by compensating the
aforementioned systematic deviations in a subsequent mapping. We have
used the data of the deviation between the desired and the actual
frame positions from 10 earlier maps for a correction that feeds ``wrong''
coordinates into the program such that the stage will actually move
to the correct position. This will not only suppress overlapping images
but also improve image quality, since the focus interpolation is carried
out assuming perfect frame positioning. In figure \ref{fig:positions}b
the same map as in \ref{fig:positions}a was acquired again, but with
stage error correction enabled. There is much less overlap in the
single images and also the inset, which shows again the x-deviations
as a gray-scale image, reveals that the positions are much more precise.
Especially the deviations between even and odd lines could be reduced
significantly. Originally, deviations of up to 61 nm from the ideal
image positions were found, whereas in the corrected case this could
be reduced to a maximum of 40 nm. \replaced{Similarly the root mean square displacement could be reduced from 36 to 18 nm between the uncorrected and the corrected case.}{Similarly the mean absolute deviation could be reduced from 26 to 11 nm between the uncorrected and the corrected case.}

\begin{figure}
\centering{}\includegraphics[width=1\textwidth]{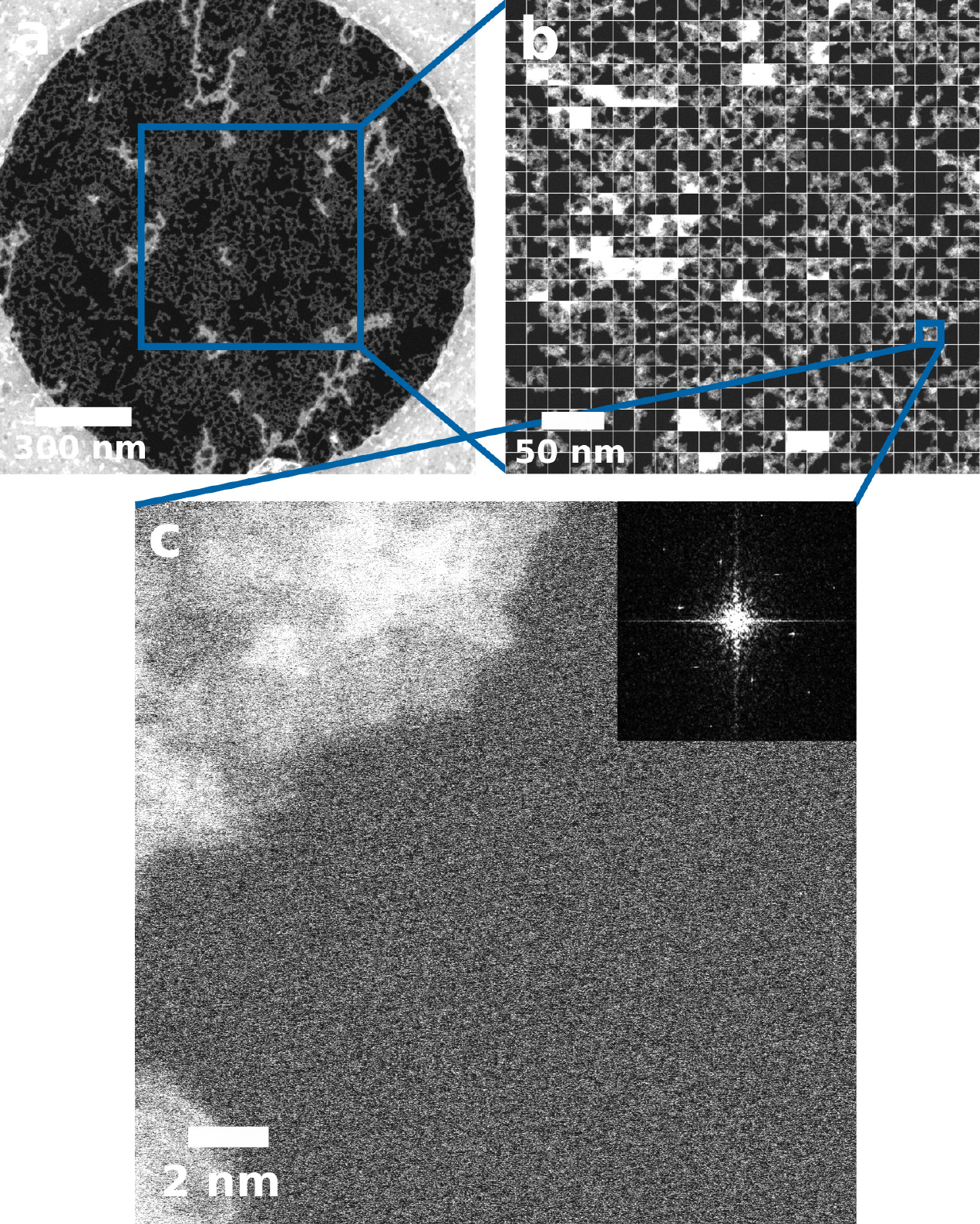}\caption{\label{fig:montage}Montage of a low-dose map consisting of 22x22
images. The total mapped area is about 0.1 \textmu m\texttwosuperior .
(a) Overview image of a hole in the Quantifoil support film covered
with monolayer graphene. (b) Montage of low-dose images acquired from
the highlighted region in (a). (c) Single low-dose image. The graphene
lattice is not directly visible, but the typical reflections are still
present in the FT.}
\end{figure}

An example where we used the software we have developed to acquire
an actual low-dose map is shown in figure \ref{fig:montage}. Although
the dose is too small to directly see the graphene lattice, it can
still be shown that the lattice information is there by analyzing
the image's FTs. In all images of the map in figure \ref{fig:montage}b
at least the first 6 FT peaks are visible, in most of them even the
first 12 which corresponds to a resolution of 2.13 Å and 1.23 Å, respectively.
Hence, the focus interpolation and stage movement can deliver atomically
resolved images also in the low-dose exposures.

\begin{figure}
\includegraphics[width=1\textwidth]{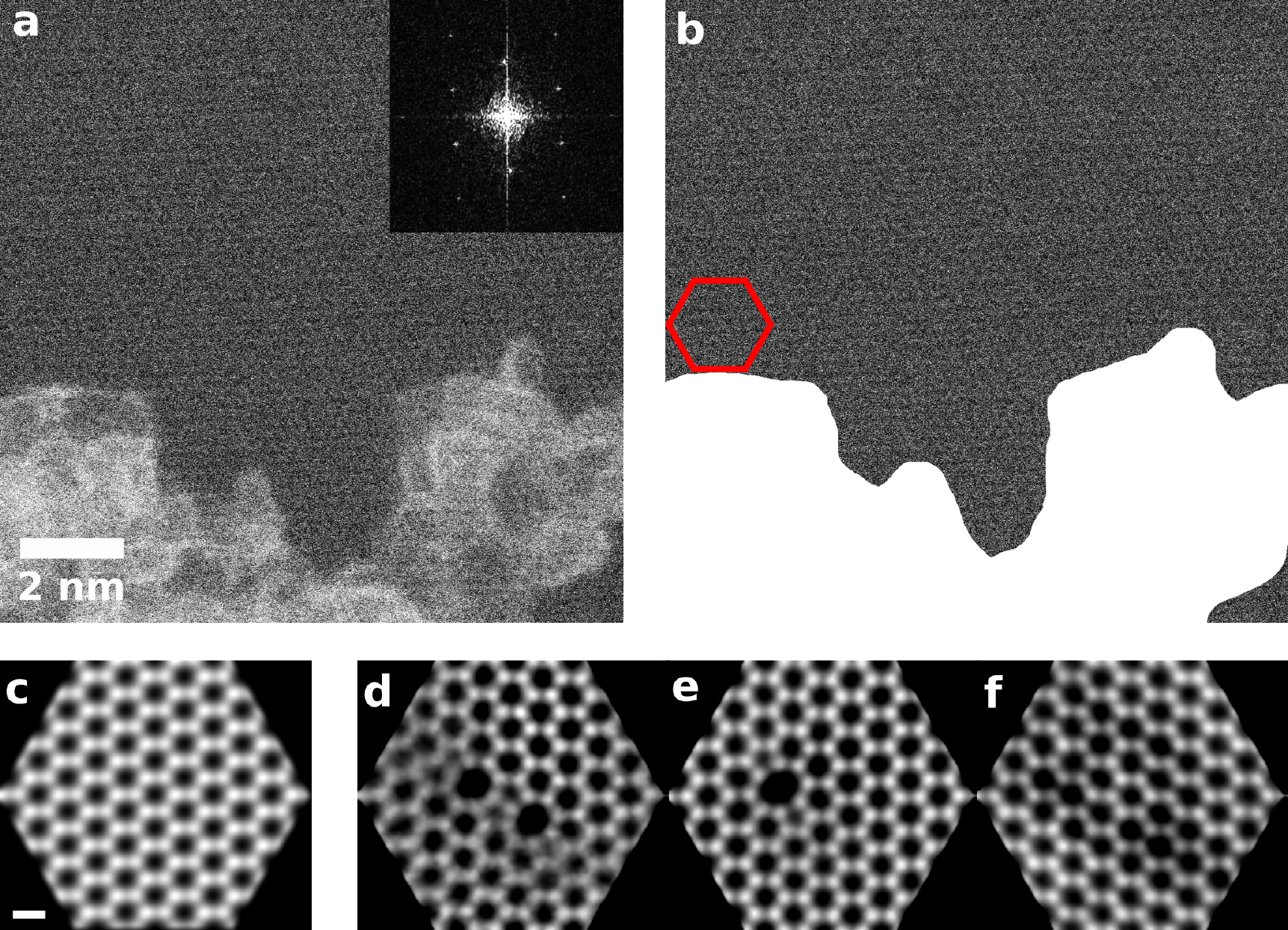}\caption{\label{fig:defect_rec}Reconstruction of point defects in graphene
from low-dose data. (a) One raw data frame. Inset: FT of the image.
Although the noise level in the image is too high to directly see
the graphene lattice, the presence of the 12 typical spots in the
FT indicates that the imaging conditions are excellent. (b) The same
image as in (a) but with the contamination masked. The red hexagon
shows how the data is cut into data cells before feeding it into the
reconstruction algorithm. (c) The sum of \textasciitilde{}90,000 data
cells shows a perfect and noise-free graphene lattice. (d-f) Result
of the reconstruction. The models are optimized in parallel and show
frequently appearing features in the raw data. The scale bar for (c-f)
is 0.25 nm.}
\end{figure}

Finally, we used the automated low-dose acquisition together with
the maxi\-mum-likeli\-hood reconstruction that we described in \cite{Meyer2014}
and \cite{Kramberger2016}. The results presented here are the first
demonstration of this new algorithm with real experimental data (Ref.
\cite{Meyer2014} used only calculations and \cite{Kramberger2016}
used experimental data that was artificially resampled to emulate
low-dose acquisition). Our test structures were defects in a graphene
membrane, which are not particularly radiation sensitive, but have
the advantage that they are well known from previous works \cite{Meyer2008e,Kotakoski2011a,Robertson2012,Kotakoski2014}.
For this experiment we used the microscope at an acceleration voltage
of 100~kV, where graphene is no longer stable and defects are created
during imaging \cite{Meyer2012}. Because the dose in single images
is too low to generate defects, we scanned up to 80 times at each
stage position in order to ensure a sufficient defect density.
The dose per exposure in this case is $1.8\cdot10^{4}e^{-}/A^{2}$
and the total accumulated dose per location up to $1.5\cdot10^{6}e^{-}/A^{2}$.
An example exposure is shown in figure \ref{fig:defect_rec}a. The
dose is too low to discern the lattice and especially any defects,
but the location and orientation of the lattice can still be obtained
from a FT of the image. Also, the contaminated region of the graphene
can still be recognized. In this data set, the contaminated areas
of the graphene sheet are now masked out, as illustrated in figure
\ref{fig:defect_rec}b. For the reconstruction, the remaining data
has to be cut into small hexagonal cells. The size of one cell is
indicated by a red hexagon in figure \ref{fig:defect_rec}b. From
the 3403 raw images recorded for this experiment, about 90,000 data
cells could be extracted by dividing the entire clean region of graphene
into slightly overlapping cells.

The starting point of the reconstruction is an empty graphene lattice
as shown in figure \ref{fig:defect_rec}c, obtained by simply averaging
all data cells. From here, the maximum likelihood reconstruction seeks
for a set of models that best describes the actually observed data.\added{ We use the following formula for calculating the likelihood of obtaining our snapshots under the assumption of the model images:}
	\begin{equation}
	\added{L=\prod_{f=1}^{F}\sum_{m=1}^{M}{\frac{w_m}{S}}\sum_{s=1}^{S}{P_{m,f,s}}}
	\end{equation}
	\begin{equation}
		\added{P_{m,f,s}=\prod_{i=1}^{I}{P(k_{f,i},\lambda_{m,s(i)})}}
	\end{equation}
\added{In this formula, $k_{f,i}$ denote the raw data values of every frame $f$ in each pixel $i$. $\lambda_{m,s(i)}$ are the expectation values of every model $m$ in each pixel $s(i)$ under a lattice symmetry operation $s$. The models are weighted with $w_m$. Every $P_{m,f,s}$ is the probability to observe a frame $f$ for a given model $m$ under a symmetry operation $s$. $P(k, \lambda)$ is the probability to observe $k$ counts for a corresponding expectation value $\lambda$. The reconstruction algorithm works with any probability density function $P(\lambda, k)$. It is, however, important to match the actual distribution function of the raw data as well as possible. We therefore use empirical histograms that can be collected from equivalent pixels for the results presented here.}
The implementation of the reconstruction \replaced{is}{was} described in \added{more }detail in
\cite{Kramberger2016}. The resulting images effectively correspond
to high dose views of all deviations from the periodic lattice that
occur sufficiently often in the data (the pristine hexagonal lattice
is also always among the set of reconstructed models, since it accounts
for the overwhelming majority of pristine graphene). From the present
dataset, two prevalent defect structures could be recovered, namely
the single 585 divacancy (DV) and a pair of aligned 585 DVs. These
two models as well as the remaining empty lattice are shown in figure
\ref{fig:defect_rec}d-f. 

It is interesting to consider why the 585 DV and also the double-585
DV were observed, while the well known ``reconstructed'' versions
of the di-vacancy in graphene (the 555777 and 55556777 \cite{Kotakoski2011a})
are absent. \replaced{These defect structures all originate from two missing carbon atoms in the hexagonal graphene lattice: Removing two adjacent atoms leads to an octagon, flanked by two pentagons (585). By beam-induced bond rotations this defect can be transformed into a circular arrangement of three pentagons and three heptagons (555777). Further bond rotations lead to a hexagon surrounded by four pentagons and four heptagons (555567777). More details on the creation, migration and transformation of defects in graphene can be found in \cite{Kotakoski2011a} and \cite{Robertson2012}}{As was described before, these reconstructed versions
form from the 585 DV via beam-induced bond rotations \cite{Kotakoski2011a,Robertson2012}}.
Hence, we can speculate that in our low-dose experiment, the irradiation
dose was high enough to form DVs in large numbers (possibly by a clustering
of mono-vacancies which are highly mobile under the beam \cite{Banhart2011}),
but not high enough to convert a significant number into their reconstructed
shapes.

\section*{Conclusions}

In summary we developed a method to automatically acquire a large
number (few hundred to a few thousand) atomically resolved images
with the Nion UltraSTEM. We find that the stage and lenses are stable
enough to obtain atomically resolved images from an area up to 1 \textmu m\textsuperscript{2}
in a time span of up to a few hours without intermediate re-tuning
by the user. The low-dose data acquisition makes it possible to obtain
atomically resolved images from sample areas that are not exposed
to the electron beam before the actual image acquisition. We expect
that this will enable the study of beam-sensitive structures such
as individual organic molecules deposited on graphene, defects in
low-dimensional materials that are too mobile under the beam for a
conventional analysis, or other ultra-thin films with embedded objects.
It may further be useful for studies that simply require larger amounts
of images than can be conveniently obtained manually, e.g. a search
for rarely occurring defects or a meaningful statistical analysis
of structural details. To our knowledge, this is the first time that
a data acquisition scheme that is commonplace in the biological single-particle
analysis was applied to a \added{non-biological} material and was achieved with sufficient
precision that individual atoms can be seen in almost all exposures.
It may provide a bridge to use methods from structural biology in
the study of radiation-sensitive materials, or vice versa.

\section*{Acknowledgements}

A.M., C.K., C.H., C.M., and J.C.M. acknowledge funding from
the European Research Council (ERC) Project No. 336453-
PICOMAT.

\clearpage{}
\bibliography{library_selected}

\end{document}